# The challenge of modeling the complexity of use for the measurement of digital maturity in education

Christine Michel[1][0000-0003-3123-913X], Laëtitia Pierrot[2][0000-0003-1701-3783], Frédéric Oru[3] and Olivier Vigneau[4]

[1] University of Poitiers, Techné UR-20297, F-86000 Poitiers, France
[2] Le Mans University, CREN UR-2661, F-72000 Le Mans, France
[3] AI4Better, F-75018, France
[4] EDIFICE, F-75017, France
christine.michel@univ-poitiers.fr ; laetitia.pierrot@univ-lemans.fr ; fred@ai4better.com ; olivier.vigneau@edifice.io

**Abstract.** Measurement of teachers' digital maturity can be useful in diagnosing technology adoption in education and in supporting teachers' professional development. However, there is a lack of easily usable tools in the existing literature. By investigating the most effective ways to measure teachers' digital maturity, our research aims to fill this gap. The focus of this article is on a data-driven approach to the assessment of digital maturity in education. Following a process similar to teaching analytics (data collection, data modeling, data understanding and visualization, and contextual understanding), we draw on diary data collected from 12949 French primary school teachers using a virtual learning environment (VLE) during the 2022-2023 school year. Specifically, the paper presents results related to modeling and visualizing data. To this end, we propose a critical analysis of two automatic classification techniques, one supervised rule-based and the other unsupervised (hierarchical clustering). In addition to providing different scales to measure, we have proposed a new approach to diagnose and visualize digital maturity. In this way, the method provides insight into the diversity and intensity of technology use, contributing to a better understanding of digital maturity and providing a practical tool for assessing day-to-day teaching practices.

## 1 Teachers’ digital maturity assessment

The recent development in educational technology has not necessarily led to a corresponding increase in teachers’ digital uses. Although there are more tools available than ever, many educators have not yet had the opportunity to fully integrate them into their teaching practices. This lack of digital maturity among teachers can hinder their ability to realize the full potential of these technologies in the classroom. Therefore, it is necessary to support teachers' digital maturity. Digital maturity is defined as an objective to be attained [23]. It manifests through intense and diversified digital uses that find their place in teaching practices [20]. Measuring teachers' digital maturity is a critical issue at various levels. Assessing digital maturity can help teachers or organizations that run initial or in-service training courses identify areas for professional development. This can lead to greater efficiency and professional fulfillment. Currently, training programs and resource design rely either on large-scale observations or structured frameworks, both guiding training needs. In Europe, DigCompEdu [20] is the most suitable framework for describing the digital skills and uses expected of teachers in their day-to-day activities. Ideally, training resources or designed actions should be varied, and organized according to the selected framework, i.e. by skill area and level of maturity, and made directly accessible and freely chosen by teachers based on their needs, interests, or level of maturity. Training resources or actions are often imposed on teachers without any real personalization [15]. Studies emphasize the importance of providing teachers with tools to observe their practices and encourage self-monitoring [7, 22]. Professional development should be encouraged through these means rather than occasional diagnoses. Digital training resources are accumulating on platforms with no uniform structure across educational areas or countries, making them difficult to access.

The absence of personalized training initiatives is primarily due to the challenge of formalizing and measuring teachers' digital maturity and uses. It is difficult for individuals to identify areas of competence that require improvement, and for organizations or platforms to recommend suitable training activities for each teacher. Our work focuses on proposing methods and tools to measure teachers' digital maturity.

Conventional methods rely on self-reported questionnaires, which provide limited and restricted perspectives on technological integration in the classroom [2]. It is difficult to capture the actual classroom context with such assessment instruments. Furthermore, the authors note the imprecision of responses, as teachers may not possess the ability to accurately assess their own skills. The limitations of these methods are associated with the declarative nature of digital uses. Digital uses are constantly evolving as users adopt available technologies and integrate them into their digital practices [15]. To overcome these limitations, we propose using data-driven methods to assess maturity based on observed uses, thanks to Teaching Analytics. This approach relies on datasets to help inform instructional processes [16]. The main advantage of this approach is that it provides quantifiable, objective data, which is an asset in the understanding of digital uses [17]. These analytics complement traditional analyses of teachers and their pedagogical practices [17]. The Teaching Analytics process is based on several phases, including data acquisition, modeling, and understanding. We followed this process to understand the complexity of digital maturity.

Two research questions underpin the following work: *RQ1. What pre-processing should be applied to the collected data to consolidate its meaning? RQ2. Furthermore, what types of analysis can be used to define the digital maturity of teachers?* To answer these research questions, we analyze different methods applied to a corpus analysis based on log activity (between September 2022 and September 2023) of 12,949 primary and secondary school teachers in the Paris academic district (France) using the VLEs One and Neo.

## 2 Data-driven teachers' digital maturity assessment

To assess teachers' digital maturity, we took a data-driven approach. According to existing studies, the data analysis lifecycle has several stages [17] (see Figure 1). Step 1 involves collecting and processing the data. Step 2 involves modeling the data to give it meaning. This step is followed by step 3, which focuses on understanding the data. Finally, step 4 aims to enhance our understanding of the data by situating it in a specific context.

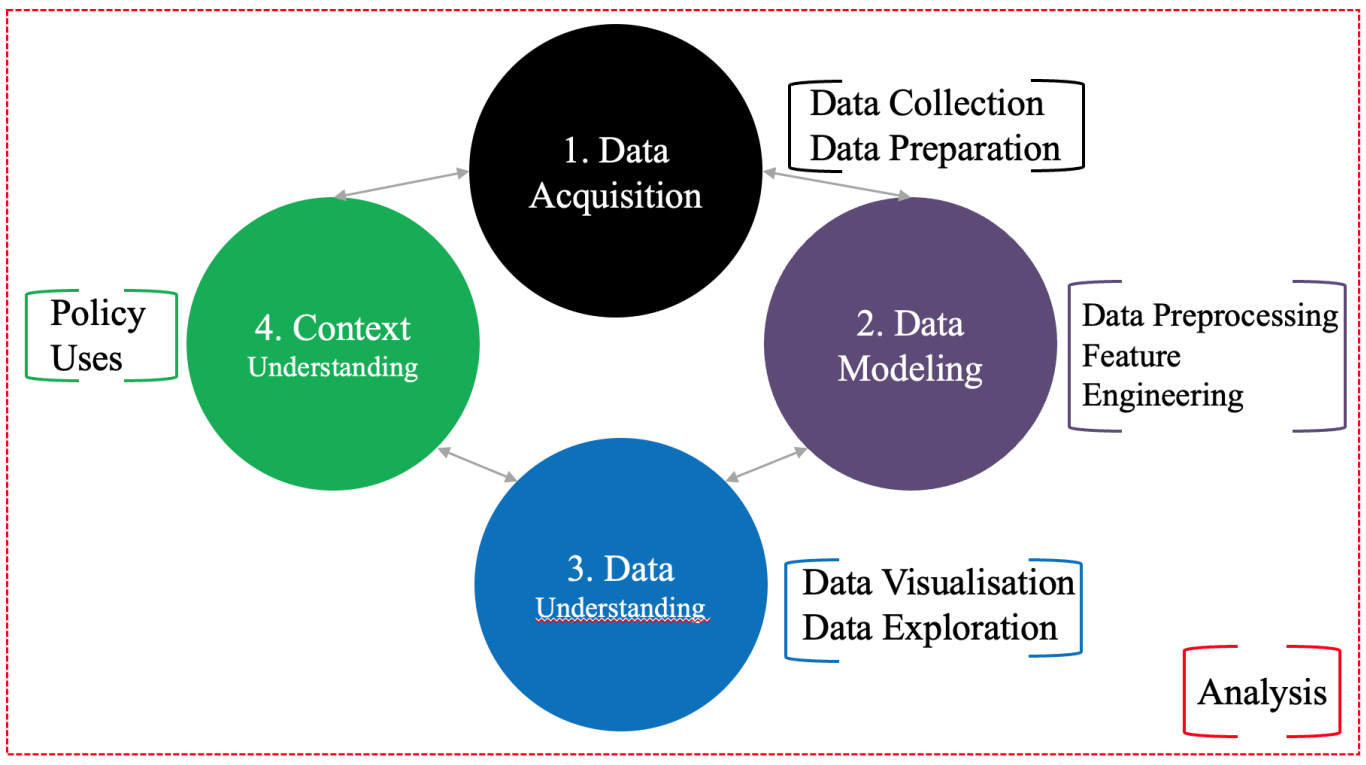

**Fig. 1.** Data analysis lifecycle adapted from [17]

**Data acquisition for digital maturity assessment.** In the educational context, data can be collected in various formats. Structured data formats, such as information stored in a database [11] have the added benefit of reporting on the dynamic use of the dynamic use of an online learning environment [25], while being captured passively. Yet, due to the wide range of data that can be collected, a preparation phase is necessary. This step allows you to remove logs of less frequent actions when analyzing logs [4].

**Data modeling for digital maturity assessment.** Data pre-processing involves ensuring that the available data can answer the question at hand [21]. It also includes transforming the data through feature engineering to generate and select meaningful variables [21]. The resulting indicators must be clear and explicit for end-users to understand [13]. Variable generation and selection can be based on the data production context [19], or by performing an analysis of variance and correlation to identify the most informative variables and eliminate duplicates [4].

Regarding the analysis of digital maturity, methods primarily rely on classification [24]. These classifications can reveal varying levels of digital maturity. Automatic classification can be approached through two main methods: supervised and unsupervised.

Supervised automatic classification involves assigning objects to predetermined classes based on specific rules. Factorial discriminant analysis was used to examine the correlation between learning strategies and participation in MOOC communities [10]. This technique can confirm whether variables are significant predictors, but it cannot consider data other than those identified at the outset during analysis. Another supervised classification technique is the decision tree, which identifies the interactions between learners and teachers in a learning environment [19]. Although this technique is useful for identifying interactions related to learning activities, its application to the teacher's activity is limited because it is difficult to define their interventions in advance.

Unsupervised classification divides numerical data into distinct groups based on the similarity between data elements. This approach does not require predefined classes since its algorithm, such as K-Means or hierarchical clustering, identifies these groups. The technique aims to divide a given set of N observations into groups (categories, classes, taxa, clusters) to group similar observations and separate dissimilar ones through clustering. These algorithms typically focus on exclusive classifications. For instance, the k-means method was used to analyze teacher behavior regarding the use of documentation services [26]. The study identified three clusters based on their use of documentation services. The k-means technique is commonly used due to its simplicity and efficiency [5]. However, it has the limitation of requiring the number of clusters to be predetermined. While K-means aims to partition the data to identify a representative data point for each cluster, hierarchical clustering seeks to create a hierarchy of clusters. From users' logs on an intelligent tutoring system (ITS), hierarchical clustering distinguishes groups based on their ITS's uses [18]. The resulting dendrogram allowed for the classification to be adjusted to four groups, retaining only the most significant variables. This method progressively groups data into a tree of clusters. Hierarchical clustering can provide richer information on the similarity structure of the data and help determine the number of clusters separately from the analysis. However, its main drawback is its high algorithmic complexity, and it is often recommended for small samples.

**Data understanding for digital maturity assessment: teaching analytics**. Applications related to Teaching Analytics focus on better understanding the teaching context, developing new training programs or designing new learning environments, evaluating teaching practices' impact on learning outcomes, or understanding teaching practices [8]. In general, understanding data contributes to teachers' reflexivity [17].

A graphical display or dashboard is the main tool used for all these actions. Its design is guided by several principles to facilitate the visualization of indicators: feedback possibilities, consideration of user preferences, and forms of support for regulation [12]. Sunburst visualization provides the ability to display data in a hierarchical, and therefore more detailed, manner. While sunbursts are inherently more complex than other graphs, their interactive nature can make them easier to read: users can zoom in on the circular area of interest, select filters, view contextual information, and navigate [3].

**Context understanding for digital maturity assessment**. Once the analyses have been performed, the context needs to be considered. In particular, understanding the context helps to take into account the constraints inherent in the context of data collec-tion. For example, the application of the GDPR in Europe is likely to limit the possibilities of using nominative data [9]. On the other hand, protocols are a rich source of data that should not be discarded. This is where mixed methods come into their own in terms of contextual resituating through triangulation techniques [26]. In addition, understanding the context is essential to making use of the analysis results and taking corrective action. At this stage, consideration of context can lead to a new cycle of data acquisition, modeling, and understanding.

**Adopted process for digital maturity assessment.** To evaluate the digital maturity of teachers, a data-driven approach provides the advantage of accurate analysis. The next sections of the paper present the steps taken in the assessment, including the acquisition of data from the VLE, verification of data quality, and modeling and understanding of the data. Unsupervised automatic classification is a promising technique for representing progressive levels of maturity and was chosen for the final part.

## 3 First steps to digital maturity assessment

### 3.1 Study context

The goal of the CoAI project is to create data-driven approaches and methods that will allow individual teachers to adopt digital resources and promote more mature, collaborative practices. One possible solution is to design a dashboard for the digital learning advisors (*délégués au numérique*) of this academy, to help them monitor VLE uses and organize in-service training activities for teachers. This collaborative effort involves various education stakeholders and is being piloted in the Paris academic district of France. The project has collected data on the actions of 12,949 primary and secondary teachers in this academy.

### 3.2 Acquisition

A virtual learning environment (VLE) is made available to all teachers, under French government guidelines [14]. The VLE is used daily by teachers and generates activity logs that can be analyzed from a teaching analytics perspective [1]. The data can be used in the context of the CoAI project of the Academic District of Paris, as the academic district authorizes the use of pseudonymized data generated by users of its VLE to meet their needs better. The processing of data is allowed within the framework of the Academy's missions in the public interest.
The data collection was structured using DigCompEdu to explore the four dimensions of domain D1 (professional engagement). Domain D1 describes four objectives for teachers: organizational communication (1.1), professional collaboration (1.2), reflexive practices (1.3), and digital continuous professional development (CPD) (1.4). To illustrate the implementation of these goals in real professional activities, four use cases were defined (see Table 1, Domain (V1)): (1) setting up communication with the class, (2) creating and selecting content, (3) managing the class and monitoring use, and (4) collaborating with peers. All services that can be used to perform these tasks have been identified, and the associated activity sensors have been verified or developed. The data collection tool allows the input of 55 variables, divided into 11, 15, 6, and 23 variables for tasks 1, 2, 3, and 4, respectively. Only 33 of these were used, as the others were not sufficiently reliable (see Table 1). Each action incremented a counter per variable and user in the log file. Using this method, we collected data from 12,949 teachers in the Paris academic district between September 1, 2022, and August 31, 2024.

**Table 1.** Variables Derived from Activity Logs.

| Domain (v1) | Domain (v2) | Actions |
|---|---|---|
| Setting up communication with the class | Logging in | 0 - I have logged in (at least once)<br>8 - I also connect from the mobile app |
| | Setting up communication with the class | 1 - I activate my account<br>4 - I upload a profile photo<br>2 - I send my first individual message<br>10 - I share a Textbook<br>50 - I publish several words (at least 2) in the Liaison Notebook<br>5 - I add a motto to my profile<br>65 - I create a forum discussion with my students<br>11 - I share several words (at least 2) in the Journal<br>51 - I insert multimedia content in the Textbook |
| Creating and selecting content | Creating and selecting content | 64 - I publish several (at least 2) blog posts<br>9 - I share a blog<br>40 - Creation of resources and sub-resources Pad, timeline, Blackboard, ...<br>35 - Creating pages (sub-resources) Multimedia notebook<br>38 - Uploading video or sound<br>14 - Library: consulting activities<br>48 - Sharing a document or folder in the Document Area<br>36 - I distribute a simple or interactive exercise or training session<br>26 - I integrate an internal recording (video, sound)<br>39 - I create a sound/video resource<br>28 - I consult resources in the Mediacenter<br>63 - I connect to the academic Moodle |
| Managing the class and monitoring uses | Managing the class and monitoring uses | 66 - Go to class settings<br>24 - Statistics: I consult class statistics<br>67 - I perform administrative actions<br>68 - Publish accounts (send email/password) |
| Collaborating with peers | Collaborating with peers | 17 - I share content (any type) with other teachers<br>15 - Library: I retrieve an activity ()<br>22 - Library: I reply to a comment ()<br>16 - Library: I comment on an activity<br>19 - Library: I publish an activity<br>21 - Library: I subscribe to a teacher |

### 3.3 Modelling process

Since uses frequencies are difficult to interpret at this stage, we decided to work with a binary data corpus (the action was performed at least once (1) or not at all (0)).

**Table 2.** Classification rules for maturity measurement.

| Rule | Conditions | Group | Subgroup |
|---|---|---|---|
| R1 | All=0 | Inactive | Inactive |
| R2 | Connex=0 & Profile =1 | Inactive | Inactive connected |
| R3 | Connex=1 | ComClass | Connection |
| R4 | R3 or (Mail or Profile =1) & (Crea & Collab & Manage =0) | ComClass | Com (Mail) |
| R5 | R4 & Other app com=1 | ComClass | Com+ |
| R6 | R3 & Blog=1 | Crea &Colab | Crea &Col (Blog) |
| R7 | R6 &Workspace=1 | Crea &Colab | Crea &Col (Blog+Workspace) |
| R8 | R6 & Other app=1 | Crea &Colab | Crea &Col (Multi app) |

Unsupervised classification methods were discarded due to their high algorithmic complexity in terms of time and memory. Rule-based classification was used to identify users according to their maturity level. The rules are described in Table 2. For example, teachers who log on at least once are in the “ComClass” group and the “connection” subgroup (see Table 2, Rule R3). If they have logged in, sent an email, or filled in their profile, but have not used the resource creation, collaboration, or management services, they are classified in the “CommClass” group and the “Com(Mail)” subgroup (Rule R4). The writing of the rules was based on the empirical results of the research on teachers' digital use. The classification resulted in 3 groups and 8 subgroups.

### 3.4 Visualization

A dashboard was created for the digital learning advisors, responsible for innovation in teaching practices, at the Paris academic district to monitor how VLE uses evolve over time and across schools (see Figure 3). VLE uses are presented as a sunburst on the left, which organizes classes into nested groups (group, subgroup, domain, activity), and a histogram on the right, which shows the number of teachers (in percentages) by domain and activity. Different modes of interaction are used to promote understanding of use and decision-making (see Figure 4). A floating thumbnail describes each area of the Sunburst as the mouse passes over it, a click on an area of the Sunburst gives details of the elements that make it up, and two filter menus are used to select the level (primary, secondary) or school, and a click on the activities shown on the right indicates the other areas concerned by this activity on the Sunburst.

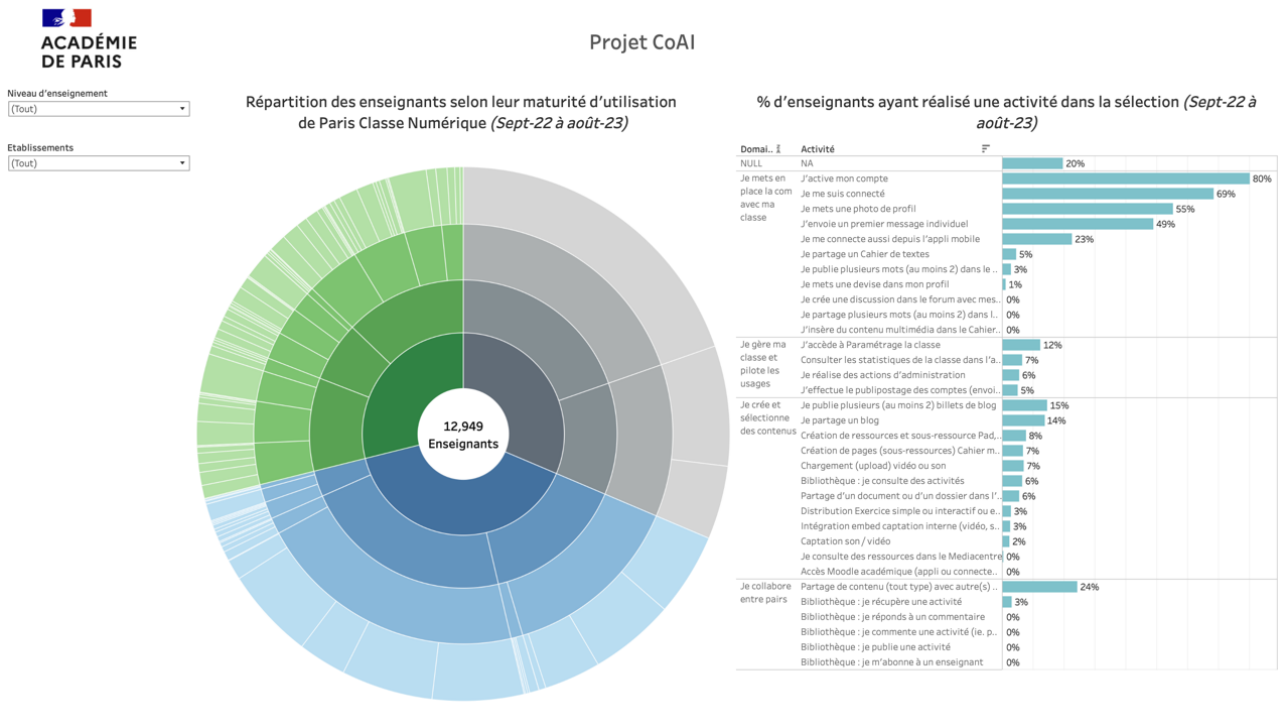

**Fig. 2.** Management dashboard for the CoAI project

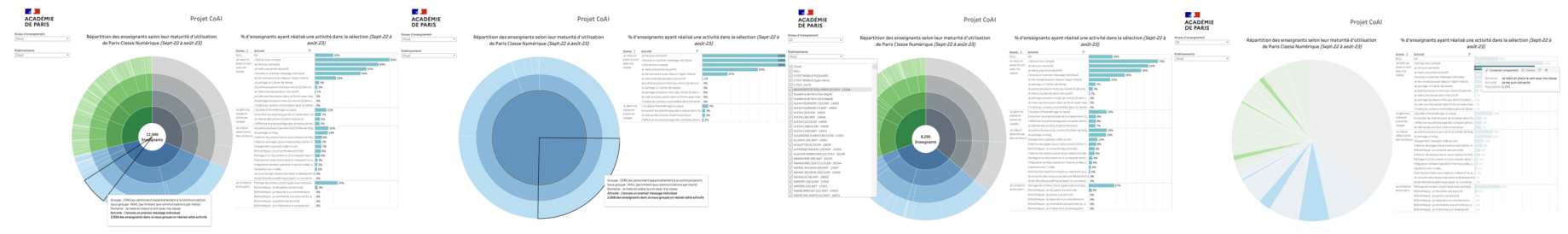

**Fig. 3.** Interaction Modes: Hover, Focus, Filter, Joint Actions

### 3.5 Data modeling challenges

This first version of the dashboard is currently operational and being tested by users in the Paris academic district. We will analyze its usability and user experience. Our critical analysis aims to improve the quality of the data analysis lifecycle, particularly in the data modeling phase.

**Data preprocessing.** Initially, we decided to recode the data and work with binary values. This allows us to represent the diversity of use and the potential for the discovery of VLE services by a user. A user who uses only 2 or 3 services is considered to be less digitally mature, while showing less diversity of uses, than a user who uses 10 services. The decision to recode the data using binary values was also justified by the complexity of characterizing the level of maturity based on the intensity of use. For example, some actions, such as filling in a profile with a photo or setting up a student account, are by their nature infrequent; they are performed once or twice a year, while others, such as sending an email, are performed almost daily. However, performing administrative actions on a platform requires a high level of maturity, whereas sending an email does not. Conversely, a user who sends an email every day has a higher level of maturity than a user who sends an email only twice a year. Managing the intensity of actions is therefore an important challenge that must be addressed if we are to measure maturity more accurately. The notion of technology appropriation integrates two notions: diversity and intensity of use.

**Classification methods.** The supervised rule-based classification method used was defined a priori, without precise knowledge of the specific behaviors of the population under consideration. Primary and secondary school teachers use VLE in different ways in their professional activities [15]. More generally, the context of the activity may play a role in the classification; data sets from the Paris academic district may not yield the same classification as those from another district. Finally, we can see that the supervised method used failed to accurately distinguish the behaviors in the sunburst; the creation and collaboration domains are merged, and the management domain is not represented in the sunburst. To evaluate the quality of the rules and possibly refine the classification, we thought it would be interesting to use an unsupervised classification method such as hierarchical clustering. The hierarchy created in hierarchical clustering allows us to test different classification levels to analyze the optimal number of classes to create and the classification criteria. We initially ruled out hierarchical clustering because of its high algorithmic complexity in terms of time and memory. Our goal is to test it on a sample of the population and derive a model of the classification criteria to be applied to the whole, to test the effectiveness of this method: to define the rules or to build other forms of modeling.

## 4 Improving the method

### 4.1 Data analysis method

Two data tables were created from the log file: the first reports the number of actions for each action/variable considered and represents the intensity of use, and the second returns 0 or 1 depending on whether an action/variable has been entered and represents the potential of a user to discover VLE services. The segmentation of the maturity according to a notion of diversity and intensity makes it possible, on the one hand, to better represent the dynamics of technological appropriation and, on the other hand, to represent, using machine learning techniques, uses that are inherently less frequent.

The variables were then grouped, for both tables, according to 5 main areas of use: logging in, organizing communication with the class, creating and selecting resources, collaborating with peers, management, and administration (see Table 1, Domain(v2)).

On each table, we performed an automatic classification using machine learning principles: hierarchical clustering. The bottom-up approach, also known as agglomerative clustering, groups the closest observations together and agglomerates them into a single cluster. This step is repeated until all clusters are grouped into a single cluster. The tree structure that groups all the clusters by hierarchical level is called a dendrogram. The dendrogram can be used to construct different classifications according to the degree of intra-class inertia (i.e., the degree of homogeneity of a class). The greater the number of classes, the lower the level of intra-class inertia and the more homogeneous the classes.

### 4.2 Maturity level clustering

**Intensity of the uses with non-supervised clustering**. Table 3 shows descriptive data for the 10 maturity levels according to use intensity (MI), obtained using a hierarchical ascending classification method. Each MI level is described according to the barycenters of the intensity of use variables grouped according to 5 domains (see Table 1, domain(V2)): (IC (Connection), ICC (Communication with class), ICS (Creation&Selection), IM (Management), ICoL (Collaboration)). Three methods were tested to order the classes according to their level of maturity: by a simple mean of the intensity of use variables by the domain (column M), by a static weighted mean (WM(stat)) which defines arbitrarily different weights according to the domains, or by a dynamic weighted mean (WM(dyn)) which adapts the weights according to the uses observed in the population. Dynamic weighting corresponds to the ratio of the maximum value of the mean of each variable divided by the mean of that variable. Static or dynamic weighting can be used to represent the complexity for a teacher to develop certain uses with greater intensity. To minimize data transformation, we sorted the MI classes from least mature (class 0) to most mature (class 9) according to the dynamic weighted mean of the uses (WM(dyn)). The NbU column represents the number of individuals in each class.

The colors (from red to green) show the evolution of the intensity of uses according to domain and MI class. Class 0 has the lowest use levels but represents the largest proportion of the teaching population (n=5305). Maturity then develops through communication (ICC) (n=783, class 1 and n=539, class 2). The most mature teachers perform all types of activity with relative regularity, in particular collaboration (ICoL) and management (IM) (classes 8, 9), but they are not that numerous (n=31 and n=7 respectively). These findings are consistent with other studies showing that teachers find it easier to develop communication uses than collaboration or management uses [2, 14].

**Table 3.** Maturity Sorted According to Use Intensity

| MI | IC | ICC | ICS | IM | ICoL | NbU | M | WM (stat) | WM (dyn) |
|---|---|---|---|---|---|---|---|---|---|
| **0** | 4.61 | 1.25 | 0.62 | 0.34 | 0.24 | 5305 | 1.41 | 0.38 | 0.40 |
| **1** | 87.55 | 19.42 | 13.32 | 5.28 | 4.44 | 783 | 26.00 | 6.95 | 7.27 |
| **2** | 211.81 | 45.43 | 21.03 | 7.17 | 6.19 | 539 | 58.33 | 11.01 | 11.74 |
| **3** | 402.46 | 74.52 | 23.39 | 12.25 | 7.62 | 340 | 104.05 | 15.33 | 16.58 |
| **4** | 625.37 | 98.37 | 34.32 | 14.64 | 8.68 | 142 | 156.28 | 20.14 | 21.97 |
| **5** | 885.47 | 121.54 | 50.07 | 16.86 | 13.76 | 104 | 217.54 | 28.03 | 30.35 |
| **6** | 1241.84 | 171.31 | 46.67 | 9.67 | 13.60 | 45 | 296.62 | 28.72 | 31.71 |
| **7** | 793.53 | 454.56 | 26.00 | 26.97 | 12.31 | 32 | 262.68 | 34.55 | 39.61 |
| **8** | 1847.29 | 242.90 | 52.71 | 59.71 | 10.29 | 31 | 442.58 | 46.55 | 51.93 |
| **9** | 3065.00 | 458.00 | 81.57 | 55.43 | 42.29 | 7 | 740.46 | 79.58 | 87.20 |
| **M** | **916.49** | **168.73** | **34.97** | **20.83** | **11.94** | | | | |
| W-dyn | 1.00 | 5.43 | 26.21 | 44.00 | 76.74 | | | | |
| W-stat | 1 | 5 | 30 | 50 | 100 | | | | |

**Diversity of the uses with non-supervised clustering.** The same calculation method was used to create the maturity table by diversity of use (see Table 4).

**Table 4.** Maturity Sorted According to Use Diversity

| DI | DC | DCC | DCS | DM | DCoL | NbU | M | WM (stat) | WM (dyn) |
|---|---|---|---|---|---|---|---|---|---|
| 0 | 0.00 | 0.00 | 0.00 | 0.00 | 0.00 | 2290 | 0.00 | 0.00 | 0.00 |
| 1 | 0.00 | 1.00 | 0.00 | 0.00 | 0.00 | 702 | 0.20 | 0.03 | 0.10 |
| 2 | 1.01 | 0.79 | 0.01 | 0.20 | 0.00 | 1060 | 0.40 | 0.08 | 0.30 |
| 3 | 1.23 | 1.80 | 0.24 | 0.15 | 0.17 | 1202 | 0.72 | 0.23 | 0.52 |
| 4 | 1.45 | 3.30 | 0.03 | 0.17 | 0.46 | 287 | 1.08 | 0.40 | 0.78 |
| 5 | 1.54 | 3.01 | 1.98 | 0.39 | 0.96 | 549 | 1.58 | 1.03 | 1.28 |
| 6 | 1.51 | 1.88 | 2.62 | 0.93 | 1.02 | 424 | 1.59 | 1.28 | 1.41 |
| 7 | 1.51 | 2.93 | 0.80 | 2.32 | 0.71 | 349 | 1.66 | 1.22 | 1.45 |
| 8 | 1.57 | 1.97 | 3.27 | 2.98 | 1.09 | 148 | 2.18 | 1.98 | 2.01 |
| 9 | 1.73 | 3.53 | 4.57 | 1.66 | 1.19 | 317 | 2.53 | 1.93 | 2.12 |
| **Mean** | 1.16 | 2.02 | 1.35 | 0.88 | 0.56 | | | | |
| W-dyn | 1.75 | 1.00 | 1.49 | 2.29 | 3.60 | | | | |
| W-stat | 1 | 5 | 30 | 50 | 100 | | | | |

**Diversity of the uses with supervised clustering.** To compare the supervised and non-supervised methods, we calculated the averages of each variable according to the classes identified with the rules, calculated the means and weighted means, and organized the maturity levels in ascending order of WM(dyn) values (see Table 5).

**Table 5.** Maturity Sorted According to Clustering Rules

| MR | Cluster | DC | DCC | DCS | DM | DCoL | NbU | M | WM (stat) | WM (dyn) |
|---|---|---|---|---|---|---|---|---|---|---|
| 0 | Start(No) | 0.00 | 0.00 | 0.00 | 0.00 | 0.00 | 1868 | 0.00 | 0.00 | 0.00 |
| 1 | Start(profil) | 0.00 | 1.12 | 0.00 | 0.00 | 0.01 | 679 | 0.23 | 0.03 | 0.09 |
| 2 | N/A(.) | 0.33 | 0.61 | 0.06 | 0.06 | 0.03 | 67 | 0.22 | 0.06 | 0.13 |
| 3 | Com(Con-nect) | 0.84 | 0.82 | 0.00 | 0.14 | 0.00 | 1280 | 0.36 | 0.07 | 0.22 |
| 4 | Com(Mail) | 1.22 | 1.93 | 0.00 | 0.33 | 0.00 | 848 | 0.70 | 0.15 | 0.40 |
| 5 | Com(Com+) | 1.38 | 2.79 | 0.00 | 0.62 | 0.77 | 261 | 1.11 | 0.66 | 0.83 |
| 6 | Crea&Col(Bl og) | 1.47 | 2.27 | 1.88 | 0.80 | 0.78 | 999 | 1.44 | 1.01 | 1.19 |
| 7 | Crea&Col(Bl og&W) | 1.37 | 2.35 | 1.80 | 0.93 | 0.77 | 350 | 1.45 | 1.02 | 1.20 |
| 8 | Crea&Col(M ulti app) | 1.36 | 2.28 | 2.27 | 1.10 | 0.85 | 976 | 1.57 | 1.19 | 1.35 |
| | M | **0.88** | **1.57** | **0.67** | **0.44** | **0.36** | | | | |
| | w-dyn | 1.78 | 1.00 | 2.36 | 3.55 | 4.42 | | | | |
| | w-stat | 1 | 5 | 30 | 50 | 100 | | | | |

## 4.3 Maturity visualization

As shown in Figure 5, the new classifications allow more precise indicators to be constructed. The 4 target use cases are represented by intensity and diversity. The two indicators, diversity, and intensity, are different but coherent, showing that these two axes of observation are useful and complementary for analyzing maturity. The diversity indicator shows that about 20% of the services are not yet used (gray area). The intensity indicator shows that 40% of the population does not use VLE. The indicators also allow a more precise comparison between schools (see Figure 6). We can see that School A is less mature than School B. To promote and stimulate the use of digital technology, it may be useful to show non-user teachers in School A the services their school uses, improving their positive image of VLE and increasing the diversity of use within the school. School B is more digitally mature. It has a greater variety and intensity of use. To increase digital maturity, it is possible to work on intensity by showing use cases different from the services already in use.

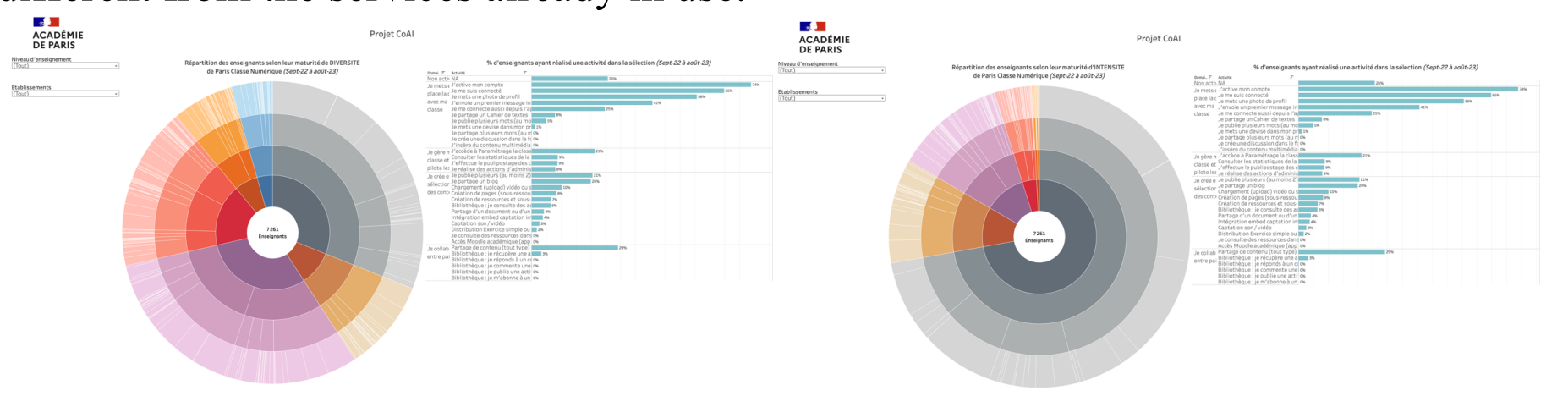

**Fig. 4.** Maturity management Dashboard: diversity (left) and intensity (right) of the primary population (n=7261)

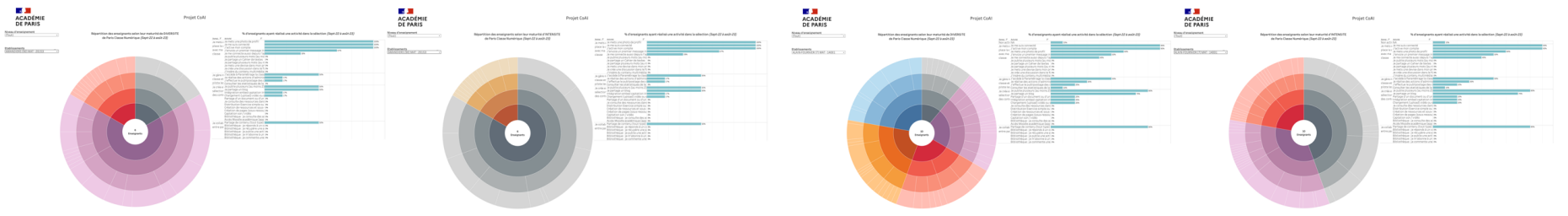

**Fig. 5.** Maturity (diversity and intensity) comparison between schools A (left) and B (right)

# 5 Conclusion and perspectives

## 5.1 Main findings

This paper addresses two main research questions: what pre-processing strategies should be applied to consolidate the meaning of a data set derived from the use of the VLE by 12,949 primary school teachers (RQ1), and what analyses should be performed to measure the level of digital maturity in education (RQ2). More broadly, this paper addresses the challenge of correctly modeling the complexity of use in a data-driven approach.

Our findings for RQ1 are directly in line with a teaching analytics perspective. Regarding data collection, the work of collecting and preparing data from the VLE was guided by a European framework, DigCompEdu. This framework has the advantage of being based on teachers' daily activities. Through critical analysis, we identified principles to follow to retain and normalize only meaningful variables across 33 tracked actions. Modeling the data according to the DigCompEdu skills repository also made it possible to identify the main areas and sub-areas of activity, and thus to consolidate the analyses by considering the teachers' professional objectives. Finally, we have built two types of data tables from the logs: (1) a binary table, composed of 0s and 1s depending on whether an action has been carried out over the period, to represent the diversity of VLE use, and (2) a table counting the number of actions carried out, to represent the intensity of use over the period.

In response to RQ2, our work in testing classification techniques has confirmed their usefulness in identifying maturity levels and, most importantly, in understanding the complexity of uses considering their intensity and diversity. In particular, the hierarchical approach has allowed us to develop sunburst-type visualizations that hold great promise for supporting teachers' growth in maturity. The supervised method has the advantage of directly controlling maturity factors and using calculation functions more suited to large-scale data collections but lacks precision for high maturity classes, characterized by innovative or niche uses. It is therefore difficult to define them a priori. Hierarchical ascending classification is more precise in describing these classes but requires additional processing to order them. Finally, the combination of supervised and unsupervised approaches seems to be a particularly interesting way to further assess digital maturity in education from a methodological point of view.

### 5.2 Perspectives

Our approach to using Teaching Analytics to help understand digital literacy is promising. However, future work is needed to fully explore teachers' use of technology and contribute to improving teaching practice. With this in mind, we identify limitations and outline possible future developments.

The data collection process has the disadvantage of excluding teachers whose practices are outside the frame. It would be interesting to start from a more comprehensive model [15]. On the other hand, the use of data from the VLE on its own is not entirely satisfactory and needs to be completed with other technologies daily.

Regarding data modeling, a more in-depth analysis of the intensity and diversity classification processes is needed to verify the measured processes and the accuracy and reliability of the two classifications for data modeling purposes. We propose to analyze the reliability and accuracy of the process on different samples, then compare the two classifications using correlations and analysis of variance to understand the factors that distinguish the two methods, generate new transformed variables to better support stakeholder decision-making, and define how to combine the two indicators into a single measure.

Regarding data understanding, the proposed visualizations for the dashboard's indicators and interaction modes need to be evaluated with the target users, the decision-makers of the Academic District of Paris, to determine their usability and effectiveness in supporting decision-making. We plan to propose a unified maturity indicator that combines the intensity and diversity of use, as well as a new interaction mode that would allow decision-makers to define the weights used to represent the complexity of certain uses (see w-stat in tables 3 and 4). This would allow them to better adapt the diagnosis to their population or field constraints.

**Acknowledgments.** This work was financed within the framework of the CoAI-DataStim project (Academy of Paris) and the Tea-TIME project (ANR-23-CE38-0012).

**Disclosure of Interests.** The authors have no competing interests to declare that are relevant to the content of this article.